\documentclass[conference]{IEEEtran}
\usepackage{fontawesome}
\usepackage{geometry}
\usepackage{pifont}
\usepackage{amssymb}
\usepackage{comment}
\usepackage{caption}
\IEEEoverridecommandlockouts
\usepackage{cite}
\usepackage[utf8]{inputenc}
\usepackage{amsmath,amssymb,amsfonts}
\usepackage{algorithmic}
\usepackage{graphicx}
\usepackage{textcomp}
\usepackage{xcolor}
\usepackage{multirow}
\usepackage{booktabs}
\usepackage{array}
\usepackage{etoolbox}
\usepackage{dblfloatfix}
\usepackage{floatpag}
\usepackage{float}

\usepackage[
   colorlinks,
    linkcolor=black,   
   citecolor=green,   
   urlcolor=black,    
    filecolor=black,   
    menucolor=black,   
    runcolor=black     
 ]{hyperref}
\begin{document}

\title{Denoising GER: A Noise-Robust Generative Error Correction with LLM for Speech Recognition}

\IEEEpubid{\makebox[\columnwidth]{\textsuperscript{\dag}Corresponding author \hfill} \hspace{\columnsep}\makebox[\columnwidth]{}}

\author{
\IEEEauthorblockN{
Yanyan Liu\textsuperscript{1,2}, 
Minqiang Xu\textsuperscript{2,\dag}, 
Yihao Chen\textsuperscript{2,3}, 
Liang He\textsuperscript{1,4}, 
Lei Fang\textsuperscript{2}, 
Sian Fang\textsuperscript{2}, 
Lin Liu\textsuperscript{2}
}
\IEEEauthorblockA{\textsuperscript{1}School of Computer Science and Technology, Xinjiang University, Urumqi, China}
\IEEEauthorblockA{\textsuperscript{2}Hefei iFly Digital Technology Co. Ltd., Hefei, China}
\IEEEauthorblockA{\textsuperscript{3}University of Science and Technology of China, Hefei, China}
\IEEEauthorblockA{\textsuperscript{4}Department of Electronic Engineering, Tsinghua University, Beijing, China}
}

\maketitle

\begin{abstract}
In recent years, large language models (LLM)
have made significant progress in the task of generation error correction (GER) for automatic speech recognition (ASR) post-processing. However, in complex noisy environments, they still face challenges such as poor adaptability and low information utilization, resulting in limited effectiveness of GER. To address these issues, this paper proposes a noise-robust multi-modal GER framework (Denoising GER). The framework enhances the model's adaptability to different noisy scenarios through a noise-adaptive acoustic encoder and optimizes the integration of multi-modal information via a heterogeneous feature
compensation dynamic fusion (HFCDF) mechanism, improving the LLM's utilization of multi-modal information. Additionally, reinforcement learning (RL) training strategies are introduced to enhance the model's predictive capabilities. Experimental results demonstrate that Denoising GER significantly
improves accuracy and robustness in noisy environments and exhibits good generalization abilities in unseen noise scenarios.

\end{abstract}
\section{Introduction}

\label{sec:intro}
Automatic speech recognition (ASR) technology has become one of the most important technologies in modern society due to its ability to efficiently and accurately transcribe audio content. With the advancement of deep learning techniques, ASR has significantly improved in terms of accuracy and efficiency. However, ASR performance is often severely affected when faced with complex background noise interference \cite{01mu2024automatic}. To address this challenge, various methods such as language model (LM) rescoring(rank)\cite{02deliberation, 03kannananalysis, 04shancomponent} and post-processing error correction\cite{05leng2021fastcorrect,06leng2021fastcorrect}are widely used in the ASR decoding process, aiming to improve the linguistic coherence and accuracy of the recognition results. In recent years, large language models (LLM) have become a key driving force in the field of artificial intelligence, thanks to their rich knowledge storage and excellent task-processing capabilities. LLM have demonstrated outstanding performance in various natural language processing tasks. The emergence of large models has provided new opportunities and possibilities for achieving breakthroughs in ASR tasks in noisy scenarios. The integration of LLM with ASR technology has become a current research hotspot. Recent studies\cite{07liao2023improving,08ma2023n-best-t5,09naderi2024towards,10chen2023hyporadise,11chen2022noiserobust} have shown that large language models can extract useful information from ASR's n-best hypotheses and perform generative error correction (GER) on transcription results. A key characteristic of GER techniques is their reliance on pretrained LLM, which benefit from the rich contextual information they provide. 

Although GER has made significant progress in improving ASR performance, it still faces many challenges in noisy environments. First, GER relies on the n-best hypotheses list generated by the speech base model as training data. Since these hypotheses lists are pre-generated and fixed, they may limit the performance and generalization ability of the GER framework. Second, existing GER methods mainly rely on language information for transcription prediction, without fully utilizing the acoustic information in the speech signal. This limitation can lead to discrepancies between the corrected results and the real speech content. 
To address the shortcomings, recent studies\cite{12chen2024uadf,13Radhakrishnan2023whisper-llama,14yu2024connecting,15hu2024listen,16mu2024mmger} have attempted to incorporate acoustic information as additional input to LLM, exploring the use of multi-modal information to further optimize GER performance in ASR post-processing.

Despite some progress through multi-modal fusion, challenges remain when dealing with complex noise interference. First, existing methods have weak adaptability to different noise environments and cannot effectively handle the impact of noise on acoustic features.
Especially when background noise is complex, the model struggles to extract high-quality speech embeddings, causing information redundancy or even interfering with the LLM's focus on useful information, thus affecting GER performance. Second, most current studies\cite{17chu2023qwen-audio,18ma-embarrassingly,19yu-connecting,20fathullah-prompting} mainly concatenate speech embeddings and n-best hypotheses in a simple manner through prompts and feed them into large models to assist the GER process. However, due to the significant modal differences between audio and text data\cite{12chen2024uadf,21wang-slm}, this simple concatenation approach may lead to information redundancy or mutual interference between multi-modal information. 
The LLM may misinterpret acoustic information as a form of language “noise”, thus affecting GER performance. This is because LLM are better at processing textual information but relatively weaker at handling acoustic information. 

To address the limitations of GER in noisy environments, this paper proposes a noise-robust multi-modal GER framework. 
This framework have three improvements: First, to enhance the model’s adaptability to specific noise environments,
we design a noise-adaptive acoustic encoder (NAAE) module. This module could extracts high-quality speech embeddings rich in semantic information, to help LLM  more accurately comprehend speech content. It also generates dynamic n-best hypotheses overcoming the limitations of fixed hyp lists. 
Second, to address the differences between speech and text modalities and improve multi-modal information utilization, we propose a heterogeneous feature compensation dynamic fusion (HFCDF) mechanism, this mechanism adjusts and compensates for multi-modal feature embeddings based on the differences between acoustic and textual modalities. It also dynamically allocates fusion weights according to each modality’s contribution to the task. 
Finally, to further enhance recognition performance in noisy environments, this paper introduces a reinforcement learning-based training strategy (RL-loss). The core idea of this strategy is to construct a reward function related to ASR evaluation metrics, incorporating the minimum word error rate (WER) as one of the training objectives, and combining it with cross-entropy loss for interpolation training. 
Our contributions can be summarized as follows:
\begin{itemize}

\item We design a NAAE that reduces noise interference and dynamically generates n-best hypotheses lists, minimizing the impact of noise on GER performance. 
\item We propose a speech-text multi-modal featre HFCDF mechanism that reduces modality differences between speech and text and improves multi-modal information utilization through dynamic fusion.

\item We introduce a RL-based training strategy RL-loss on that optimizes training objectives using ASR evaluation metrics, enhancing the model’s ability to recognize noisy speech.

\item  Experiments show that the proposed Denoising GER method improves error correction robustness against noise. It achieves significant performance gains on synthesized noisy speech and demonstrates generalization to unseen noise scenariosand clean speech.
\end{itemize}

\begin{figure*}[!t]
    \clearpage
    \centering
   \includegraphics[width=\linewidth] {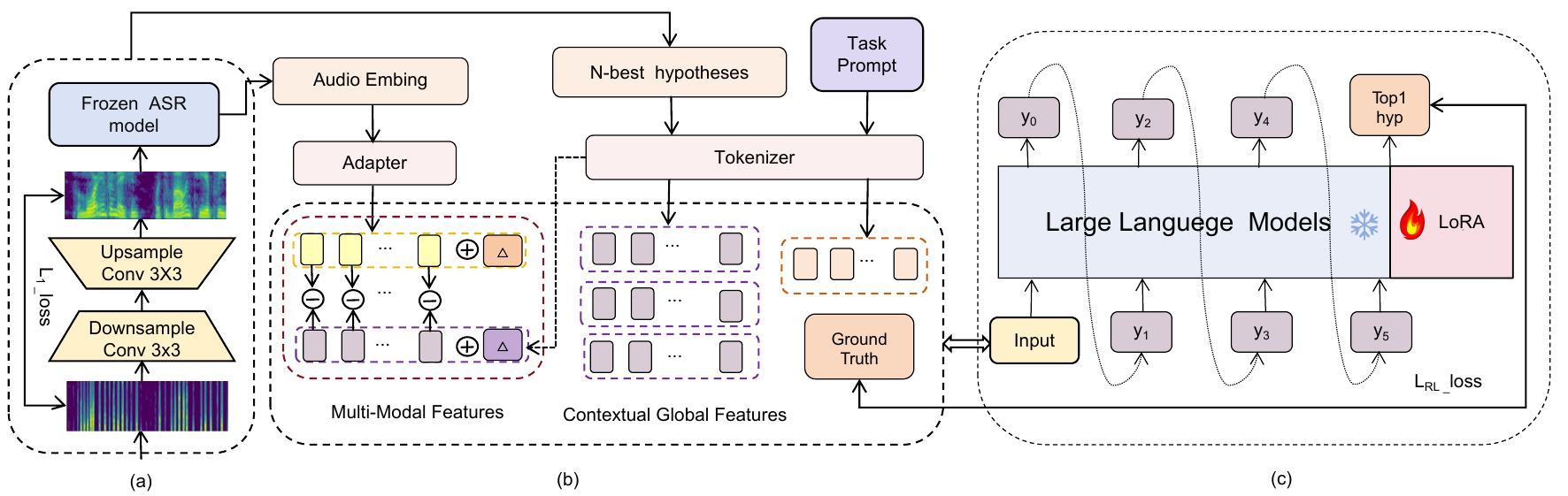} 
    \caption{the overall framework of our proposed Denoising GER with LLM for ASR, Part (a) of Figure 1 represents the adaptive process for noisy speech; Part (b) illustrates the HFC-based  multimodal feature dynamic fusion containing multimodal feature, and together with n-best hypotheses list containing global context information, forms the input for GER; Part (c) shows the RL-based fine-tuning and inference process of  the GER with  LLM.}
    \label{fig:Architecture}
\end{figure*}

\section{Related Work}
\label{sec:format}

\subsection{Exploration of LLM in Multi-modal Tasks}

In recent years, LLM based on the transformer architecture, such as ChatGPT\cite{lund2023chatGPT}, GPT-4\cite{achiam2023GPT4}, and LLama\cite{35-touvron-llama}, have attracted significant research interest due to their excellent language generation and reasoning abilities. With large model parameters and extensive training data, LLM can deeply understand linguistic structure and semantic information in text data. They have demonstrated outstanding performance in various natural language processing tasks. 
Recently, researchers have begun exploring the potential of LLM in multi-modal tasks. They aim to integrate other modalities\cite{29-li2023blip,30-lyu-macaw,31-wu-decoderonly,32-barrault-seamless,33-rubenstein-audiopalm},
into LLM to advance downstream tasks. These tasks include automatic speech recognition (ASR)\cite{13Radhakrishnan2023whisper-llama,15hu2024listen,18ma-embarrassingly}, speech translation\cite{32-barrault-seamless,33-rubenstein-audiopalm}, and audio event detection and understanding.

\subsection{Applications of LLM in Speech Recognition}

The emergence of LLM has created new opportunities for breakthroughs in ASR. This is because LLM, with expanded data and model parameters, possess strong language pattern understanding and prediction abilities. Currently, research on integrating LLM with ASR systems falls into two main categories. The first approach combines LLM with pretrained ASR models to replace traditional n-gram or neural network language models. By feeding ASR-generated text into LLM as prompts\cite{08ma2023n-best-t5,10chen2023hyporadise}, this method improves speech recognition error correction\cite{23ma-can}. For example, \cite{07liao2023improving} proposed an ASR post-processing technique that extracts semantic information from ASR transcriptions to generate more readable text. Studies\cite{08ma2023n-best-t5,09naderi2024towards} utilized LLM’ contextual learning ability to rescore ASR outputs. In\cite{10chen2023hyporadise}, researchers introduced an LLM-based ASR post-processing GER benchmark. This method fine-tunes LLM using text-only n-best hypotheses sequences generated by ASR, allowing the model to learn the mapping between n-best hypotheses and the true transcription. It significantly outperforms traditional n-gram and neural network language model (LM) rescoring (rank) methods. Additionally, to enhance the noise robustness of GER, RobustGER\cite{11chen2022noiserobust} applied knowledge distillation. This method extracts noise information from source speech and incorporates it into language embeddings for GER, achieving noise reduction in the language space. As a result, it demonstrates excellent noise robustness.

\subsection{Advances in LLM-based Multi-modal GER Frameworks}


In recent years, several representative studies have made significant progress under this paradigm. For example, SALMONN\cite{19yu-connecting} uses the Whisper model \cite{26radford-robust}{} to extract semantic content and BEATs\cite{27chen2022beats} to capture audio event information, enabling robust perception of human speech, music, and audio events. Qwen-Audio\cite{17chu2023qwen-audio} employs Whisper\cite{27chen2022beats} as an encoder and enhances performance in multi-audio tasks by designing task-specific instructions. SpeechGPT\cite{28-D-Zhang-speechGPT} unifies tokenizer for speech and text, directly incorporating source speech into the GER paradigm's n-best hypotheses list to handle multi-modal audio tasks. SLAM-ASR\cite{18ma-embarrassingly} utilizes linear layers as adapter networks for modality conversion and feature dimension reduction, achieving state-of-the-art (SOTA) performance on the Librispeech\cite{38-panayotov-librispeech} 960-hour English ASR task. MMGER\cite{16mu2024mmger} adopts a Conformer-based acoustic feature extractor and improves multi-modal speech error correction by concatenating acoustic and text embeddings using prompts.

\section{Methodology}

The framework (as shown in Figure\ref{fig:Architecture}) consists of an acoustic encoder, an adapter network, and a frozen LLM. The acoustic encoder is used to reduce noise interference, extract optimized acoustic embeddings, and generate dynamic n-best hypotheses lists from the speech signal. After the acoustic embeddings are aligned in dimensions and converted through the adapter network, they are subject to heterogeneous feature compensation and multi-modal feature dynamic fusion with the 1-best hypothesis, generating a frame-level multi-modal representation of the speech signal. 
Building on this, the LLM combines the global  information from the n-best hypotheses and leverages its powerful understanding and generative capabilities to perform fine-grained frame-level multi-modal correction and global context correction on the characters, aiming to improve ASR accuracy in complex noisy environments.

\subsection{Overview of  Generative Error Correction}

The task of ASR aims to predict its textual transcription consisting of $T$ sequential tokens $Y_{T}=\{y_{1}, y_{2},..., y_{T}\}$ with neural network. We follow the original GER benchmark, where, given a speech signal $X \in \mathbb{R}^{l}$, the pretrained ASR model is used to transcribe it into an N-best hypotheses list $\widehat{y}_{n}=\{Y_{1}, Y_{2},..., Y_{N}\}$ through beam search decoding. Each hypothesis can be viewed as an independent text representation of the speech signal. The goal of GER is to learn a hypotheses-to-transcription mapping (M$_{H2T}$) in an autoregressive manner. This can be expressed as:
\begin{equation}
Y = M_{H2T}(\widehat{y}_{n}, \theta_{l}) 
\label{eq:111}
\end{equation}
\begin{equation}
P(Y|\widehat{y}_{n}, \theta_{l}) = \prod_{t=0}^{T} P(y_{t}|Y_{\ll t}, \widehat{y}_{n}, \theta_{l}) 
\label{eq:222}
\end{equation}
where, $\theta$ represents the parameters of the learnable adapter network within the LLM, and $y_{t}$ represents the historical sequence $\{y_{1}, y_{2}, ... , y_{t-1}\}$.

\subsection{Noise-Adaptive Acoustic Encoder}

Existing LLM-based ASR research mostly focuses on optimizing the adapter network, but there is insufficient research on acoustic encoders. 
For example, SLM\cite{21wang-slm} only trains the adapter network while keeping other modules unchanged, and in\cite{20fathullah-prompting}, LoRA is used only within the LLM network. Overall, existing work either fine-tunes the encoder completely, which may lead to catastrophic forgetting, or keeps it frozen, making it difficult to fully exploit its potential on the given data. Additionally, the high computational and storage requirements for practical deployment pose challenges.
To address this, we propose a NAAE. 
Drawing inspiration from the success of efficient fine-tuning techniques in the NLP field, we integrate a small U-Net adapter into a frozen encoder as a learnable noise adaptation module. This module improves noise handling through efficient fine-tuning while preserving the encoder's original recognition performance. The adapter network takes Mel spectrograms as input and outputs adapted Mel spectrograms. Its structure consists of three downsampling layers and three up-sampling layers. 
The core function of the adapter is to act as a noise-adaptive denoising model, enhancing the acoustic model's robustness against noise. Given a noisy speech signal $X_{in}$, the adapter network yields an adaptive speech signal $X_{in}$. The adaptive process of the speech signal can be described as follows:
\begin{equation}
X_{in} = f_{\theta}(X_{in}) = X + adapter_{\theta}(X_{in}) \label{eq:adaptation_process}
\end{equation}

After training, the adapter network can receive speech signals from various noisy environments and reconstruct them back to the state recognized by the ASR model.
In the Denoising GER framework, the ASR encoder and decoder update a small portion of model parameters, and provide LLM with high-quality acoustic embeddings and a dynamically changing n-best hypotheses list. This enriches the error distribution space of the n-best hypotheses, helping the LLM learn the GER process more effectively. This process can be described as follows:
\begin{equation}
\left \{\begin{matrix} 
X_{audio} = Encoder_{ASR}(X_{in}) \\ 
Y_{hyp} = Decoder_{ASR}(X_{audio})
\end{matrix} \right. \label{eq:asr_process}
\end{equation}
The loss of the ASR acoustic encoder can be expressed as:
\begin{equation}
L_{ASR} = \min_{\theta} \lambda L_{CE}(X_{in}^{'}, Y) + (1 - \lambda) L_{1}(X_{in}^{'}, X_{in}) \label{eq:asr_loss}
\end{equation}
where $X_{in}$ represents the FBank features, and $Y_{hyp}$ is the n-best hypotheses prediction result by the ASR decoder.

\subsection{multi-modal Speech-to-Text GER Based on LLM}

Under background noise interference, the effectiveness of GER faces two main challenges: First, noise interference leads to significant errors in the initial transcription generated by the ASR system, making the subsequent correction task more complex and difficult. Second, most existing traditional GER methods rely on a single text modality, lacking the support of acoustic information which can lead to correction errors or over-correction, limiting the model's robustness and accuracy. 
To address these issues, this paper proposes a noise-robust multi-modal error correction method, inspired by MMGER\cite{16mu2024mmger}. 
By incorporating speech embeddings as additional input to LLM, this approach fully utilizes the multi-modal information processing from both speech and text, thereby improving the accuracy and fidelity of corrections in noisy environments. 

Specifically, given a speech signal, the ASR model generates acoustic embeddings $X_{audio}$ and the n-best hypothesis sequence $Y_{hyp}$. The LLM's tokenizer is then used to process the n-best hypotheses into character-level embeddings $Y_{tok}$. Next, an adapter network aligns the speech embeddings $X_{tok}$ with the characterlevel 1-best hypothesis and acoustic embeddings $X_{audio}$. Subsequently, the multi-modal representation of the speech is concatenated with the n-best candidate transcriptions along the feature dimension and used as input to the large model, generating a fine-grained multi-modal representation. At the same time, by incorporating a discourselevel n-best hypotheses list with global contextual information, the model performs a collaborative optimization of frame-level multi-modal correction and global context correction, improving the accuracy and reliability of generative error correction. This process can be described as:
\begin{equation}
\left \{\begin{matrix} 
X_{tok} = Adapter(X_{audio}) \\ 
Y_{tok} = Tokenizer_{LLM}(Y_{hyp})
\end{matrix} \right. \label{eq:multi-modal_process} 
\end{equation}
Therefore, equations \eqref{eq:111} and \eqref{eq:222}  should be revised as:
\begin{equation}
Y = M_{H2T}(X_{tok}, Y_{tok}, \theta_{i}) \label{eq:revised_7}
\end{equation}
\begin{equation}
P(Y|\widehat{y}_{n}, X, \theta_{i}) = \prod_{t=0}^{T} (y_{t} | Y_{<t}, X_{tok}, Y_{tok}, \theta_{l}) \label{eq:revised_8}
\end{equation}
\subsection{Multi-modal Heterogeneous Feature Compensation  Dynamic Fusion Mechanism}

In early research, speech embeddings and text embeddings were typically aligned through direct concatenation\cite{16mu2024mmger,17chu2023qwen-audio} or cross-attention\cite{13Radhakrishnan2023whisper-llama} before being fed into the LLM decoder for inference and prediction. However, since speech and text come from different modalities, there is a large modality gap between them. 
Simple concatenation methods may cause LLM to fail to effectively understand the acoustic embeddings.  
This is because LLM are adept at processing text embeddings, but they are less familiar with processing acoustic embeddings.
To address this issue, inspired by \cite{12chen2024uadf,21wang-slm,36-zhang2023mrcn,37-zhang-multi}, we propose a HFCDF, and we calculate the vector difference between the speech embedding and the text embedding, then use this difference as a compensation vector to perform cross-modal compensation with the original acoustic and text features. This establishes a spatial correspondence between the two different modality features, enabling the model to more effectively balance the understanding and fusion of speech and text features. 
This process can be described as follows:  
\begin{equation}
\left \{\begin{matrix} 
\triangle_{x}=X_{tok}-Y_{tok} \\ 
\triangle_{y}=Y_{tok}-X_{tok}
\end{matrix} \right. \label{eq:delta_xy}
\end{equation}
\begin{equation}
\left \{\begin{matrix} 
X'_{tok}=X_{tok}+k\triangle _{y} \\ 
Y'_{tok}=Y_{tok}+(1-k)\triangle _{x}
\end{matrix} \right. \label{eq:adjusted_tokens}
\end{equation}
Where, $X_{tok}$ represents the acoustic embedding, $Y_{tok}$ represents the text embedding, and $\triangle_{x}$ and $\triangle_{y}$ represent the vector differences between speech and text modalities, $k$ is a hyperparameter.

Furthermore, to further optimize feature fusion, we measure the correlation between each modality's features and the target output by calculating the cosine similarity. 
The core idea is that when multiple modalities express similar information or contribute consistently to the task, they provide more assistance to the task decision, and thus should be assigned higher weights. Specifically, we calculate the similarity scores between the acoustic modality, the text modality, and the target output, and use the Softmax function to convert these scores into weights. This allows us to dynamically adjust the fusion weights for each modality. This process can be described as:
\begin{equation}
\left \{\begin{matrix} 
R_{a}=Cosine(X'_{tok}, y) \\ 
R_{t}=Cosine(Y'_{tok}, y)
\end{matrix} \right. \label{eq:similarity_scores}
\end{equation}
\begin{equation}
\mu =\frac{e^{R_{a}}}{e^{R_{a}}+e^{R_{t}}} \label{eq:dynamic_weight}
\end{equation}
\begin{equation}
X'_{mmc}=Concat(\mu X'_{tok}, (1-\mu )Y'_{tok-top1}) \label{eq:dynamic_fusion}
\end{equation}
\begin{equation}
P(Y)=\prod _{t=0}^{T}P(y_{t}|Concat(X'_{mmc}, Y'_{tok}), \theta _{1}) \label{eq:final_prediction}
\end{equation}
Here, $y$ represents the target output, and $R_{a}$ and $R_{t}$ are represent the similarity scores between the acoustic modality and the text modality with the target output. Compared to traditional concatenation or static weighting methods, this dynamic weighted fusion method based on modality similarity better handles the complementarity between modalities, improving the quality of multi-modal information fusion. As a result, it enhances the performance of GER while avoiding interference from redundant information.

\subsection{The global learning of  Denoising GER }

WER is an important metric for evaluating the performance of ASR systems. However, since LLM are typically trained using cross-entropy loss, which does not align with the evaluation metric used during inference, this mismatch can lead to suboptimal training results. 
Specifically, cross-entropy loss  optimizes the model's output probability distribution at the character level, but in noisy conditions, the blurred or missing information in the speech signal can amplify the ASR model's prediction errors at the character level. Does not directly reflect the impact of these WER. Moreover, cross-entropy loss focuses more on local information and lacks global semantic understanding at the sentence level. In environments with strong background noise, this limitation makes it difficult for the model to optimize higher-level linguistic information, leading to more compound errors.

To address this issue, given the outstanding performance of RL in sequence modeling tasks, we propose a RL-loss. By constructing a reward function based on ASR metrics, we use the minimum Word Error Rate (WER) as one of the training objectives and combine it with cross-entropy loss for interpolated training\cite{bai2024seed-asr}. By directly optimizing the WER metric, the model can better focus on the ultimate performance goal of the speech recognition task. The cross-entropy loss ensures basic prediction performance through character-level supervision. The combination of both makes the model more robust in noisy environments, optimizing global error rates while improving character-level correction effects. The training objective of the RL-loss can be expressed as:
\begin{equation}
\mathcal{L}_{RL}=\frac{1}{N}\sum _{y_{i}\in n-best(x, N)}\widehat{p}(y_{i}|x)(w(y_{i}, y^{*})-\bar{w}) \label{eq:reinforcement_learning_loss}
\end{equation}
Here, $w(y_{i}, y^{*})$ represents the Word Error Rate (WER) between the ground-truth and each hypothesis $y_{i}$ in the n-best list. $\bar{W}$ denotes the average WER of the n-best hypotheses, and $\widehat{p}(y_{i}|x)$ represents the normalized likelihood probability of the predicted hypothesis. Finally, we combine the speech-text joint multi-modal representation and the n-best hypotheses containing global contextual language information as the input to the LLM. By leveraging the LLM's powerful text understanding and generative capabilities, we perform multi-modal correction and context correction on the noisy speech ASR predictions. The total loss of this framework consists of the $\mathcal{L}_{LLM}$, $\mathcal{L}_{ASR}$, and $\mathcal{L}_{RL}$. It can be expressed as follows:
\begin{equation}
\mathcal{L}_{Denoising\space GER}=\mathcal{L}_{LLM}+\alpha \mathcal{L}_{ASR}+\beta \mathcal{L}_{RL} \label{eq:total_loss}
\end{equation}
where $\alpha$ and $\beta$ are adjustable hyperparameters that allow flexible adjustment of the contribution from each part during training.

\section{Experimental Setup}

\subsection{Dataset}

To evaluate the performance of the proposed method, we conducted experiments using both synthetic and real noisy speech corpora. For the synthetic speech, we used the 100-hour training subset (train-clean-100) from LibriSpeech\cite{38-panayotov-librispeech} and randomly mixed it with 20,000 noise segments (55 hours in total) extracted from the DNS-Challenge dataset at signal-to-noise ratios (SNR) $\in \{5, 6, 7,..., 20\}$ dB to generate the training dataset. The test sets were generated by randomly mixing the same conditions with the test-clean and test-other subsets. Additionally, we evaluated the model's generalization ability to unseen noisy domains by mixing the test set with the MUSAN\cite{39-snyder-musan} corpus at the same SNR settings to create out-of-domain noisy data for testing in other noisy speech scenarios.

For real-world noisy speech, we performed experiments on the CHINE4 challenge data. A six-channel distant microphone array and a close-talk microphone were used for recording by having people read text prompts from the Wall Street Journal (WSJ0) corpus\cite{41-paul1992design}. The CHIME4\cite{40-menne-chime4} dataset contains two types of noisy speech: real and simulated noisy speech. The real data was recorded under four challenging noisy environments: bus, cafe, pedestrian area, and street junction. The simulated data are generated by mixing clean utterances with background noise recorded in the four environments. This study selected the dev-real and test-real parts of the one-channel track real section from CHIME4 as the test set to evaluate the performance of the method under real noise conditions. Finally, to evaluate the model's generalization and stability, we also tested it on the clean LibriSpeech test sets, test-clean and test-other, to assess its performance on clean data. This helps verify whether the framework can maintain recognition accuracy for clean speech while improving noise robustness.

\subsection{Model settings}

For the proposed noise-robust speech-text multi-modal Denoising GER framework based on LLM, we use the open-source Whisper-Large v2\cite{26radford-robust}and Qwen-7B \cite{42-bai2023qwen} (7B parameters) from the Hugging Face website as the front-end acoustic encoder and frozen LLM decoder, respectively. The hidden dimensions of the encoder and LLM are 1024 and 4096, respectively. Therefore, the input and output channels of the adapter network are 1024 and 4096, with 8 subsampling steps. For efficient fine-tuning of the LLM, we adopt the popular low-rank adapter (LoRA) \cite{24hu-lora} tuning strategy, where the rank r is set to 8, and LoRA is added to the query, key, value, and output layers of each transformer block, and the beam size was set to 5. The number of trainable parameters is only 26M, which accounts for just 0.32\% of the LLM’s total parameters. During fine-tuning, we use the Adam optimizer with a learning rate set to 2e-4 and a warmup step of 100. The number of training epochs is set to 5, with a batch size of 32 with nvidia A40 46GB GPUs. The maximum input sequence length is set to 1024. The hyperparameters  is set to 0.7, and $\alpha$ and $\beta$ are both set to 0.2. 

\section{Experimental Results and Analysis}

\begin{table*}[h]
\centering
      {\fontsize{8}{12}\selectfont  
\renewcommand{\arraystretch}{1}  
\setlength{\parskip}{0pt}  
    \centering
\caption{Inference results of our proposed Denoising GER framework compared to existing methods on different background noise datasets, as well as the inference results on other unseen background noise data and clean data without added noise(WER\%).}
\begin{tabular}{>{\centering\arraybackslash}m{0.5cm}>{\centering\arraybackslash}m{2.4cm}>{\centering\arraybackslash}m{1.4cm}>
{\centering\arraybackslash}m{1.4cm}>
{\centering\arraybackslash}m{1.2cm}>{\centering\arraybackslash}m{1.2cm}>
{\centering\arraybackslash}m{1.4cm}>
{\centering\arraybackslash}m{1.4cm}>
{\centering\arraybackslash}m{1.4cm}>{\centering\arraybackslash}m{1.4cm}}
 \toprule
\multicolumn{1}{c}{\multirow{3}{*}{\textbf{ID}}} & \multicolumn{1}{c}{\multirow{3}{*}{\textbf{Method}}}
& \multicolumn{4}{c}{\textbf{In-domain Data}} 
& \multicolumn{4}{c}{\textbf{Out-of-domain Data}} \\
\cmidrule(lr){3-6} \cmidrule(lr){7-10}
& & \multicolumn{2}{c}{\textbf{DNS\_LS100h}} & \multicolumn{2}{c}{\textbf{CHIME4}} & \multicolumn{2}{c}{\textbf{MUSAN\_LS100h }} &  \multicolumn{2}{c}{\textbf{Clean\_LS100h }} \\
\cmidrule(lr){3-4} \cmidrule(lr){5-6} \cmidrule(lr){7-8} \cmidrule(lr){9-10} 
& & test-clean & test-other & dev\_real & test\_real & test-clean & test-other & test-clean & test-other \\
\midrule
\multirow{4}{*}{A} 
& ASR baseline        & 7.61 & 11.85 & 6.08 & 7.83 & 6.65 & 10.82 & 2.72 & 5.66 \\
& LLM$_{Rank}$\cite{07liao2023improving}  & 6.65 & 8.51 & 5.63 & 6.97 & 6.60 & 9.13 & 2.86 & 5.62 \\
& GER~\cite{10chen2023hyporadise}                  & 6.32 & 8.87 & 5.03 & 6.62 & 6.42 & 10.27 & 2.53 & 5.18 \\
& robustGER~\cite{11chen2022noiserobust}      & 6.07 & 8.66 & 4.86 & 6.34 & 6.37 & 9.76 & 2.47 & 5.14 \\
\midrule
\multirow{3}{*}{B} 
& Qwen-Audio~\cite{17chu2023qwen-audio}     & 6.12 & 8.49 & 4.72 & 5.75 & 6.45 & 8.72 & 2.14 & 4.59 \\
& MMGER~\cite{16mu2024mmger}              & 6.07 & 8.32 & 4.53 & 5.23 & 6.44 & 8.55 & 2.35 & 4.63 \\
& \textbf{ours}  & \textbf{5.82} & \textbf{8.24} & \textbf{4.28} & \textbf{5.32} & \textbf{6.18} & \textbf{8.16} & \textbf{2.08} & \textbf{4.32} \\
\bottomrule
\end{tabular}

    \label{tab:mainresult}}
\end{table*}

\subsection{Main Result}

 We conducted experiments on in-domain datasets noisy LS100h and CHIME4, as well as out-of-domain datasets MUSAN\_LS100h and Clean\_LS100h. The results are shown in Table \ref{tab:mainresult}, with the best results highlighted in bold. 
 Groups A and B represent the existing unimodal GER methods, and current mainstream multi-modal GER methods, respectively. 
 1) Overall, our Denoising GER frameworks achieved competitive WER on both in-domain and out-of-domain datasets. 
 It outperformed traditional LM rescoring, 
unimodal GER, and mainstream multi-modal GER methods. 2) On the in-domain datasets DNS\_LS100h and CHIME4, our method performed well in various noisy environments. 
This shows that our model can handle different noise scenarios and has strong robustness and error correction capabilities. 3) On the out-of-domain dataset MUSAN\_LS100h, our method achieved WERs of 6.18\% on test-clean and 8.16\% on test-other. These results are significantly better than those of traditional GER and multi-modal GER methods. This indicates that Denoising GER maintains high recognition accuracy in unknown noisy conditions, demonstrating good generalization ability. 4) On the clean dataset , These results are clearly better than those of the ASR baseline and LM rank methods, and show a competitive advantage compared to other GER methods. Notably, improving performance in noisy conditions did not negatively affect recognition accuracy on clean speech,proving that the model maintains versatility and stability across different scenarios. In summary, our proposed Denoising GER framework shows stable performance on both in-domain and out-of-domain datasets. Our method achieves leading results on datasets with known noise, unknown noise, and clean speech. This demonstrates that the proposed framework fully utilizes complementary acoustic and textual multi-modal information, achieving higher speech recognition accuracy in complex noisy environments while maintaining good generalization ability and robustness.

\subsection{Ablation Study}
To verify the adaptive capability of NAAE module and its impact on the GER performance,
We conducted the following ablation experiments (see Table \ref{tab:NAAE} for details). The results show: 1) The fine-tuning approach of the acoustic model significantly affects GER performance. For example, comparing the “Frozen” and “Full Ft” models, the frozen acoustic model (Frozen) generally results in higher WER after correction. This indicates that fine-tuning the acoustic model appropriately can improve GER performance. 2) Although full fine-tuning performs better on some test sets, it requires a larger number of training parameters (1571.5M) and may lead to catastrophic forgetting, which affects the model’s original recognition ability and adaptability to new task scenarios. In contrast, the LoRA fine-tuning method preserves the original capabilities while improving efficiency and avoiding catastrophic forgetting. 3) The NAAE proposed by us combines the advantages of the LoRA fine-tuning and its original recognition performance, and has obvious advantages in both the learnable parameter amount of the model and the denoising performance of the model.
\begin{table}[h]
    \centering
      {\fontsize{8}{12}\selectfont  
\renewcommand{\arraystretch}{1}  
\setlength{\parskip}{0pt}  
     \caption{Other configurations and network structures are consistent with Qwen-Audio, exploring the impact of different fine-tuning methods of the acoustic model on GER performance.}
     \begin{tabular}{>{\centering\arraybackslash}m{0.8cm}>{\centering\arraybackslash}m{0.8cm}>{\centering\arraybackslash}m{1.2cm}>{\centering\arraybackslash}m{1.2cm}>{\centering\arraybackslash}m{1.1cm}>{\centering\arraybackslash}m{1.1cm}}
        \toprule
         \multirow{2}{*} {\parbox[t]{0.8cm}{\textbf{\centering Audio \\ Encder }}}  & \multirow{2}{*} {\parbox[t]{0.8cm}{\textbf{\centering Trianed\\Param.}}} & \multicolumn{2}{c}{\textbf{DNS\_LS100h}} & \multicolumn{2}{c}{\textbf{CHIME4}} \\
        \cmidrule(lr){3-4} \cmidrule(lr){5-6}
        & & test-clean & test-other & dev-real & test-real \\
        \midrule
        Frozen & 21.5M & 7.16 & 10.69 & 5.92 & 8.12 \\
        Full Ft & 1571.5M & 6.35 & 9.24 & 5.01 & 6.58 \\
       \midrule
        LoRA & 41.5M & 6.22 & 8.82 & 4.89 & 6.32 \\
            \textbf{NAAE} & 26.5M & \textbf{6.05} & \textbf{8.58} & \textbf{4.77} & \textbf{6.21} \\
        \bottomrule
    \end{tabular}
    \label{tab:NAAE}}
\end{table}


\begin{table}[h]
    \centering
    {\fontsize{8}{12}\selectfont  
\renewcommand{\arraystretch}{1}  
\setlength{\parskip}{0pt}  
    \caption{The ASR model remains frozen, and the LLM is fine-tuned using Lora. We aim to explore the impact of different multi-modal feature fusion methods on the GER performance.}
    \begin{tabular}{>{\centering\arraybackslash}m{2cm}>{\centering\arraybackslash}m{1.2cm}>{\centering\arraybackslash}m{1.2cm}>{\centering\arraybackslash}m{1.1cm}>{\centering\arraybackslash}m{1.1cm}}
        \toprule
         \multirow{2}{*} {\parbox[t]{2cm}{\textbf{\centering Multi-Modal \\ 
     \ \ \ \  \ \ Fusion}}} & \multicolumn{2}{c}{\textbf{DNS\_LS100h}} & \multicolumn{2}{c}{\textbf{CHIME4}} \\
        \cmidrule(lr){2-3} \cmidrule(lr){4-5}
        & test-clean & test-other & dev-real & test-real \\
        \midrule
       
        Linguistic-Only  & 6.48 & 9.39 & 5.23 & 6.42 \\
        Acoustic-Only  & 7.16 & 9.88 & 5.86 & 6.93 \\
       \midrule
        Add   & 6.41 & 8.93 & 4.90 & 6.25 \\
        Concat  & 6.35 & 8.81 & 4.78 & 6.12 \\
        Transformer  & 6.28 & 8.77 & 4.65 & 6.13 \\
        \textbf{HFCDF}  & \textbf{6.17} & \textbf{8.61} & \textbf{4.53} & \textbf{6.08} \\
        \bottomrule
    \end{tabular}
    \label{tab:MMF}}
\end{table}


\begin{table*}[h]
    \centering
      {\fontsize{8}{12}\selectfont  
\renewcommand{\arraystretch}{1}  
     \caption{The contribution of each component of our proposed noise-robust Denoising GER framework to the error correction effect is compared with WER (\%) on real and synthetic speech recognition datasets.}
       \begin{tabular}{>{\centering\arraybackslash}m{1.8cm}>{\centering\arraybackslash}m{1.8cm}>{\centering\arraybackslash}m{1.8cm}>
       {\centering\arraybackslash}m{1.8cm}>
      {\centering\arraybackslash}m{1.8cm}>
      {\centering\arraybackslash}m{1.8cm}>{\centering\arraybackslash}m{1.8cm}}
             \toprule

            \multicolumn{1}{c}{\multirow{2}{*} {\textbf{NAAE}}} & \multicolumn{1}{c}{\multirow{2}{*} {\textbf{HFCDF}}} & 
            \multicolumn{1}{c}{\multirow{2}{*} {\textbf{RL-loss}}} &\multicolumn{2}{c} {\textbf{DNS\_LS100h}} &  \multicolumn{2}{c}{\textbf{CHIME4}} \\ 
             \cmidrule(lr){4-5} \cmidrule(lr){6-7}
           & & & \textbf{test-clean} & \textbf{test-other} & \textbf{dev-real} & \textbf{test-real} \\ 
          \midrule
           $\times$ & $\times$ & $\times$ & 6.32 & 8.87 & 5.03 & 6.62 \\  
         \checkmark& $\times$ &$\times$ & 6.05 & 8.58 & 4.77 & 6.21 \\
          $\times$ & \checkmark &$\times$  & 6.17 & 8.61 & 4.53 & 6.08 \\ 
         $\times$ &$\times$ &  \checkmark  & 5.92 & 8.72 & 4.58 & 4.92 \\
          \checkmark &  \checkmark & $\times$ & 5.87 & 8.49 & 4.35 & 4.75 \\
          \checkmark &  \checkmark &  \checkmark  & \ \textbf{5.82}  & \textbf{8.24} & \textbf{4.28} & \textbf{5.32} \\
            \toprule
        \end{tabular}
    \label{tab:444} }
\end{table*}



In Table \ref{tab:MMF}, we present the results of comparison experiments between our proposed HFCDF method and other fusion approaches. The results show: 1) Combining multi-modal information significantly improves speech error correction in noisy environments compared to using only text information or only acoustic information. This suggests that multi-modal fusion provides more useful information, enhancing the model’s GER ability. 2) Using text information for single-modal GER achieves better performance than using acoustic information. This indicates that, for speech recognition error correction, the text modality provides more stable and useful information than the acoustic modality. LLM are better at handling text information, and especially when speech quality is low or noisy, LLM may treat low-quality speech embeddings as noise, negatively impacting GER performance. 3) Our proposed HFCDF method achieves the best results. Compared to other fusion methods like Add, Concat, and Transformer, HFCDF outperforms them on all test sets. 
This indicates that our method more effectively models the complementary relationship between speech and text, alleviating cross-modal discrepancies and improving overall GER performance.

To verify the contribution of each component in the Denoising GER, we conducted various ablation experiments on both in-domain and out-of-domain datasets. The experimental results are shown in Table \ref{tab:444}. From the table we can see that adding only the NAAE model leads to significant improvement in GER, demonstrating the strong robustness of the NAAE in handling different noisy scenarios. When using the HFCDF mechanism alone for speech-text multi-modal feature fusion, this module effectively reduces the differences between modalities by adjusting and compensating cross-modal information. Additionally, the introduction of HFCDF mechanism enhances the complementarity of multi-modal features, laying the foundation for subsequent performance improvement. Furthermore, combining the NAAE module with the HFCDF mechanism results in even greater performance gains. The NAAE module provides higher-quality adaptive acoustic representations for the LLM, creating a more favorable condition for the HFCDF module to compensate and optimize multi-modal features. This combination enables the model to minimize the differences between modalities and reduce the interference of cross-modal information, further enhancing overall performance. We validated the effectiveness of adding the RL-loss. These experimental results confirm the effectiveness of each component in our Denoising GER method.

\subsection{Error Analysis}
Table \ref{tab:5555} presents a case study to validate the effectiveness of the proposed Denoising GER framework. The ASR model’s predictions include common substitution and insertion errors. From the table, we can observe: 1) Our Denoising GER method, after correction, matches the Ground Truth perfectly, achieving the best results to date. This shows that the Denoising GER framework has a significant advantage in handling noise interference and error correction. In contrast, existing LLM-based GER [10] and RobustGER [11] frameworks, while showing some improvement, still suffer from semantic bias. This highlights Denoising GER’s excellent performance in noise robustness and finegrained error correction. 2) Although the current best multi-modal GER methods, such as UADF [12] and MMGER [14], combine speech-text multi-modal information, their correction performance is limited when dealing with complex errors caused by noise interference. Especially in noisy environments, some speech signals are masked by noise, leading to misrecognition of phonetically similar words. Traditional methods fail to effectively counteract noise  interference, resulting in weak alignment between acoustic embeddings  and semantics. In comparison, Denoising GER reconstructs clear speech signals through noise-adaptive acoustic models and combines heterogeneous compensation with a multi-modal dynamic fusion strategy, significantly improving correction accuracy.
\begin{table}[h]
    \centering
     {\fontsize{8}{12}\selectfont  
\renewcommand{\arraystretch}{1}  
\setlength{\parskip}{0pt}  
    \centering
     \caption{Error Case Analysis. We compare the GER results of our proposed Denoising GER framework with existing LLM-based GER frameworks. }
\begin{tabular}{m{2.1cm}
<{\centering\arraybackslash}m{4.8cm}<{\centering\arraybackslash}m{0.7cm}<{\centering\arraybackslash}}
 \specialrule{1.2pt}{0pt}{0pt}
\textbf{Method} & \textbf{Utterance} &\textbf{WER(\%)} \\
\midrule
ASR baseline & pour \textcolor{red}{may raise}  over all \textcolor{red}{chille at}  serve & 57.14 \\
LM$_{Rank}$\cite{07liao2023improving} & pour \textcolor{red}{may raise} over all \textcolor{red}{chile at} serve & 57.14 \\
GER\cite{10chen2023hyporadise} & pour mayomaise over \textcolor{red}{chiles} and  \textcolor{red}{served}  & 42.86 \\
RobustGER\cite{11chen2022noiserobust} & pour mayomaise over  \textcolor{red}{chilies} and serve & 14.29 \\
Qwen-Audio\cite{17chu2023qwen-audio} & pour mayomaise over all \textcolor{red}{chill} and \textcolor{red}{served} & 28.57 \\
MMGER\cite{16mu2024mmger} & pour mayonnaise over all \textcolor{red}{chili} and serve & 14.29 \\
Ours & pour mayonnaise over all chill and serve & 0 \\
\midrule
Ground Truth & pour mayonnaise over all chill and serve & - \\
 \specialrule{1.2pt}{0pt}{0pt}
\end{tabular}
\label{tab:5555}  }
\end{table}

\section{CONCLUsION}

This paper proposes a noise-robust multi-modal error correction framework Denoising GER to improve the error correction ability of LLM in noisy speech. This framework combines the advantages of a noise-adaptive acoustic encoder, a multi-modal feature dynamic fusion mechanism, and a reinforcement learning training strategy, effectively reducing the interference of background noise, optimizing the extraction of speech features and the fusion effect of multi-modal features, enabling the LLM to better understand the speech content and improving the LLM's utilization of multi-modal information. Additionally, the RL training strategy further improves the model's adaptability to noisy speech, making the error correction effect more in line with practical application requirements. Experimental results validate the effectiveness of our framework in various noisy environments, with strong generalization ability in unseen noise scenarios, offering an effective solution for speech recognition in complex noisy conditions. Future research can further explore lighter acoustic models and more optimized multi-modal fusion strategies on this basis to further enhance the noise robustness and overall performance of the model.
\bibliographystyle{ieeetr}  
\bibliography{IEEEabrv, references}  

\end{document}